\documentclass[letterpaper]{article}
\pdfoutput=1 

\usepackage{jheppub} 

\usepackage{graphicx}
\usepackage{comment}

\usepackage{natbib}

\title{
Searching for cosmic strings via black hole spin-down  \\
}

\author[a]{Sameer Ahmed,}
\author[a]{Michael Kavic,}
\author[b]{Steven L. Liebling,}
\author[a]{Matthew Lippert,}
\author[a]{Mohammad Mian,}
\author[c]{and John Simonetti}

\affiliation[a]{Department of Chemistry and Physics, \\SUNY Old Westbury,\\Old Westbury, NY, United States}

\affiliation[b]{Department of Physics, \\Long Island University,\\Brookville, NY, United States}

\affiliation[c]{Department of Physics, Virginia Tech, Blacksburg, VA, United States}

\emailAdd{sahmed35@oldwestbury.edu}
\emailAdd{kavicm@oldwestbury.edu}
\emailAdd{Steve.Liebling@liu.edu}
\emailAdd{lippertm@oldwestbury.edu}
\emailAdd{mmian3@oldwestbury.edu}
\emailAdd{jhs@vt.edu}

\abstract{
Cosmic strings that are attached to rapidly spinning black holes can extract significant amounts of rotational energy and angular momentum. Here, we study the effect on primordial black holes, which are expected to form with one or more cosmic strings attached.  Although large primordial black holes of mass greater than $\mathcal{O}(10 M_\odot)$ are predicted to rapidly spin up due to accretion soon after forming, we argue that cosmic strings will spin them down again. We show that if there are cosmic strings with tension greater than $10^{-20}$, and, provided at least one remains attached, the spins of large primordial black holes should consequently be observed to be near zero. We also investigate the effect on a supermassive black hole of capturing a cosmic string and discuss the possibility of observing the subsequent spin down by its effect on a pulsar orbiting the black hole.}

\preprint{OW-2024-2}

\begin{document}

\maketitle
\flushbottom

\section{Introduction}
\label{sec:intro}

Cosmic strings are conjectural remnants of phase transitions in the early universe and represent an intriguing astronomical manifestation of high-energy physics.  They are one-dimensional topological defects naturally appearing in many particle physics models, including many grand unified theories (see \cite{Vilenkin:2000jqa, Hindmarsh:1994re, Sakellariadou:2009ev} for reviews). In the context of string theory, fundamental strings and D-strings can also manifest as cosmic superstrings \cite{Polchinski:2004ia}.

Efforts to detect cosmic strings have exploited several possible phenomena.  Oscillating strings produce a stochastic gravitational wave background \cite{Vachaspati:1984gt}. Cosmic strings can also develop cusps and kinks which emit strong bursts of gravitational radiation \cite{Damour:2000wa, Damour:2001bk, Damour:2004kw}. The amplitude of this gravitational wave signal depends primarily on the string's mass density, or tension, $\mu$.\footnote{We work in units where $\mu$ is dimensionless. Restoring the units, the tension is given by $G\mu /c^2$.}

Perhaps most exciting are the recent results from the North American Nanohertz Observatory for Gravitational Waves (NANOGrav), looking for ultra-low-frequency gravitational waves by observing millisecond pulsars. NANOGrav has found evidence of a stochastic gravitational wave background compatible with a network of cosmic strings in the range $10^{-12} < \mu < 10^{-10}$  \cite{NANOGrav:2023hvm, Ellis:2023tsl}. Although not a firm detection, the observations offer tantalizing evidence for the existence of cosmic strings in the early universe. 

Direct searches by the LIGO/Virgo/KAGRA (LVK) gravitational wave (GW) observatory network for burst signals have set an upper bound of $\mu < 3 \times 10^{-7}$, and analysis of the stochastic GW background yields a tighter constraint of $\mu < 10^{-10}$ \cite{LIGOScientific:2021nrg}. A stochastic background would also induce characteristic noise in the timing of pulsar signals.  Results from the Parkes Pulsar Timing Array currently give the strongest bound: $\mu < 1.5 \times 10^{-11}$ \cite{Lasky:2015lej, Blanco-Pillado:2017rnf}.  
Microlensing of extragalactic sources by strings could probe down to tensions as small as $\mu = 10^{-13}$ \cite{Chernoff:2019pjr}.  Future GW detectors, such as LISA, could potentially be sensitive to $\mu = 5.8 \times 10^{-18}$ \cite{Blanco-Pillado:2017rnf}.

Here, we investigate the interaction between a cosmic string (CS) and a black hole (BH), with a view toward observable signatures.  A CS colliding with a BH will be gravitationally captured, with the black hole ending up stuck like a bead on a string \cite{DeVilliers:1997nk, Snajdr:2002aa, Dubath:2006vs}.  A short CS loop might be expected to quickly fall in to the BH, but a very long CS, in particular one extending beyond the cosmological horizon, could persist attached to the BH for a long time.

The behavior of a CS attached to a rotating black hole has been studied in \cite{Frolov:1996xw, Vilenkin:2018zol, Deng:2023cwh}.  Interestingly, the cosmic string can extract rotational energy and angular momentum \cite{Kinoshita:2016lqd, Igata:2018kry} from the BH by a generalization of the Penrose Process \cite{Penrose:1971} using one-dimensional relativistic strings rather than particles.  The well-known Blandford-Znajek mechanism \cite{Blandford:1977ds}, in which a force-free electromagnetic field extracts rotational energy from an accreting BH, is also an example of this stringy Penrose Process but with magnetic flux lines acting as the strings \cite{Kinoshita:2017mio}.  Unlike the Blandford-Znajek mechanism, in which the rotational energy powers highly visible jets emanating from accreting black holes, the energy extracted by cosmic strings is not easily observable. However, the spin down and energy loss of the BH is potentially detectable.

An important question to consider is how likely is it that a given BH will, in fact, collide with and capture a CS. In \cite{Xing_2021}, the rate for a black hole to capture cosmic strings was estimated.  A stellar-mass BH presents a small target, and the chance of capturing a CS is extremely low; the expected time to capture a single cosmic string would be far longer than the age of the universe. However, for a supermassive black hole, such as Sagittarius A* (Sgr A*), and for very low cosmic string tension $\mu \lesssim 10^{-18}$, there is an order-one probability for one or more collisions during the black hole's lifetime.

Observing the spin down of a supermassive black hole due to cosmic strings, although challenging, is possible, at least in principle.  We propose a technique by which the energy lost by the BH could be detected by its effect on a pulsar orbiting the BH.  However, in addition to finding an appropriate pulsar sufficiently close to the BH, the precision of pulsar timing measurements would need to improve significantly to detect mass loss due to phenomenologically viable cosmic strings with $\mu < 10^{-11}$.

A particularly interesting alternative scenario is the interaction of a cosmic string with a primordial black hole (PBH) \cite{Vilenkin:2018zol}. PBHs may have been created by large density fluctuations in the early universe with a very wide range of possible masses, in principle, down to the Planck scale (see \cite{Sasaki:2018dmp, Green:2020jor, Villanueva-Domingo:2021spv, Escriva:2022duf} for recent reviews). The possibility that large black holes recently observed by LVK may have primordial origin \cite{Bird:2016dcv} has spurred renewed interest. Moreover, recent observation by the James Webb Space Telescope (JWST) may have discovered evidence for the existence of primordial black holes \cite{Matteri:2025vnv, Zhang:2025tgm, 2025arXiv250821748J}.

In the early universe, at the time of PBH formation, the cosmic string network would be sufficiently dense that there will be, on average, one or more cosmic strings in a typical over-dense region, which collapses to form a PBH.  The PBH is then created with one or more cosmic strings already attached \cite{Vilenkin:2018zol}.

A typical PBH formed by over-densities during the radiation-dominated era is not expected to form with significant spin \cite{Mirbabayi:2019uph, DeLuca:2019buf}.  However, there are mechanisms by which PBHs form with large spins, e.g. \cite{Cotner:2017tir, Harada:2017fjm}. Even if the PBH is formed with low spin, subsequent accretion of gas and dark matter can spin it up.  In particular, it was shown in \cite{DeLuca:2020bjf} that this accretion process is mostly completed early in the history of the universe and that large PBHs are expected to have higher accretion rates, resulting in higher spins.  As a result, a PBH with a mass greater than about $30M_\odot$ is predicted to have near-extremal spin when later observed.

However, if the PBH is formed with one or more long, horizon-crossing cosmic strings attached, then after being spun up by accretion, the PBH would subsequently be spun down by those strings.  As a result, later observations would find high-mass PBHs with very small angular momentum, in contrast to the prediction of \cite{DeLuca:2020bjf}.  

An important caveat is that cosmic strings can become detached from the black hole.
Two long strings attached to a black hole can intersect and reconnect, leaving one long string disconnected from the black hole and a short half-loop, which may rapidly fall into the horizon. 
In such a case, no string would remain to spin the black hole down.
However, the dynamics of cosmic strings and their respective networks depend on detailed properties, perhaps best explored with numerical simulations. One example is Ref.~\cite{Deng:2023cwh} which showed that short string loops attached to a rotating black hole, rather than falling in to the horizon, can elongate, growing away from the black hole, and providing an alternative mechanism to extract rotational energy.

Accurate measurements of BH spins have become increasingly possible as techniques have improved from gravitational wave observations of binary BH mergers and from X-ray emission spectroscopy of BH accretion disks (for a review, see \cite{Reynolds:2020jwt}).  With a sufficiently large dataset of black hole spin observations, it will become possible to distinguish, at least statistically, those PBHs formed in the early universe from other BHs formed later, for example, via stellar collapse. Observations of highly spinning PBHs would then rule out, or at least highly constrain, the existence of cosmic strings at the time of PBH formation.

The outline of the remainder of this paper is as follows: In section \ref{sec:energy_extraction}, we review the mechanism by which cosmic strings remove energy from black holes.  Then, in section \ref{sec:PBH}, we argue that cosmic strings would have effectively spun down PBHs being observed in the present epoch.  In section \ref{sec:supermassiveBH}, we discuss the possibility of observing the cosmic string-induced spin down of a supermassive black hole via pulsar timing measurements.  Finally, we discuss open questions and future directions in section \ref{sec:discussion}.

\section{Energy extraction by cosmic strings}
\label{sec:energy_extraction}

The Penrose Process \cite{Penrose:1971} is a mechanism in classical general relativity by which energy can be removed from a rotating black hole.  An object that enters within the ergosphere of the black hole will necessarily co-rotate with the black hole.  The object is then split into two parts, and one part launched in the backward (contra-rotating) direction.  This piece will still be co-rotating, albeit slower, and can have negative energy and angular momentum as measured by an asymptotic observer.  When this negative energy piece ultimately falls through the event horizon, the energy and angular momentum of the black hole are reduced.  The remaining part of the object receives a kick and can escape the black hole with greater energy and angular momentum than it had originally.

The available energy is limited by the spin of the black hole. A Kerr black hole\footnote{A Kerr BH with angular momentum $J$ and mass $M$ has spin parameter $a=J/M$, an event horizon radius $r_h = M + \sqrt{M^2 - a^2}$, and an angular velocity $\Omega_h = \frac{a}{a^2+M^2}$. The radius of the ergosphere is $r_\mathrm{ergo} = M + \sqrt{M^2 - a^2 \cos^2\theta}$ where $\theta$ is the azimuthal angle in Boyer-Lindquist coordinates. The dimensionless Kerr parameter is defined as $\chi = a/M$ and $\chi = 1$ for an extremal Kerr BH.}
with angular momentum $J$ has a mass $M$ given by
\begin{equation}
    M = \sqrt{M_\mathrm{irr}^2 + \frac{J^2}{4M_\mathrm{irr}^2}} \ ,
\end{equation}
where the irreducible mass $M_\mathrm{irr}$ is given by the horizon area: $M_\mathrm{irr}^2 = \frac{A}{16\pi}$.  The Penrose process can continue extracting energy until $J=0$ and the black hole is completely spun down, while the area theorem ensures that $M_\mathrm{irr}$ does not decrease.  At this point all of the rotational energy $E_\mathrm{rot} = M-M_\mathrm{irr}$ has been removed, leaving a non-rotating BH with $M = M_\mathrm{irr}$. As a result, at most $29\%$ of the original BH energy may be extracted.

The Penrose process is actually a quite general description of classical energy extraction from a rotating black hole, with a wide range of physical scenarios. Superradiance (for a review, see~\cite{Brito:2015oca}) is one such example.  Another such process involves an externally sourced magnetic field threading the black hole, extracting rotational energy via the Blandford-Znajek~(BZ) process~\cite{Blandford:1977ds}. The equivalence of the BZ process and energy extraction by a protruding CS is demonstrated in~\cite{Kinoshita:2017mio,Igata:2018kry}.

In particular, a long, relativistic CS with one end entering the ergosphere is dragged around by the black hole.  If the CS is rigidly rotating with an angular velocity
$\omega$ that is less than the angular velocity $\Omega_h$ of the BH, i.e., if $\omega < \Omega_h$, then there is an outward radial flow of angular momentum and energy along the string \cite{Kinoshita:2016lqd}. In the limit of a slowly rotating black hole, $J < M^2$, and averaging over the angle at which the string emerges from the ergosphere, the outward angular momentum flux $q$ is
\begin{equation}
    q = 2M^2(\Omega_h - \omega) \, .
\end{equation}

For a CS with tension $\mu$, the energy extraction rate is $dE/dt =  \mu q\omega $.  Restoring the dimensionful constants, this can be written as \cite{Kinoshita:2016lqd}:
\begin{equation}
\label{eq:energy_extration_rate}
\frac{dE}{dt} = \left(4.5 \times 10^{58} \frac{\textrm{erg}}{s}\right)  \mu \left(\frac{q/a}{1/2}\right)\left(\frac{\omega/\Omega_h}{1/2}\right) u(a) \ ,
\end{equation}
where the dimensionless function $u(a) = \left(1-\sqrt{1-a^2/M^2}\right)$. When $a=0$ the BH is no longer spinning and $u(0) = 0$, which shuts off the energy extraction.

As the black hole loses rotational energy to the string, its total mass decreases. For typical values of $a$, $q$, and $\omega$, the final three factors of \eqref{eq:energy_extration_rate} are order one. The rate at which the BH is losing mass, assuming $a > 0$ and $\omega < \Omega_h$, is then approximately
\begin{equation}
\label{eq:mass_loss_rate}
\frac{dM}{dt} \approx - \left( 10^{34} \frac{\textrm{kg}}{s}\right) \mu \approx - \left( 10^{4} \frac{M_{\odot}}{\rm s}\right) \mu \, 
\end{equation}
which agrees with the results of \cite{Xing_2021}.

This scenario, with a slowly and rigidly rotating string of $\omega < \Omega_h$, represents a quasi-steady, late-time configuration.  The initial capture of a CS, for example, involves complicated dynamics, whose quantitative treatment is beyond this approximation and is best addressed with the numerical methods of Refs.~\cite{Xing_2021, Deng:2023cwh}.  Relative motion of the CS and the BH could result in the string rotating faster than the BH, $\omega > \Omega_h$, injecting angular momentum and spinning up the BH.  However, the angular momentum added by the string would be parametrically small due to its low tension: A segment of string with length and impact parameter both of order of the horizon radius carries angular momentum of order $\mu M^2$ compared to the black hole's angular momentum $J \sim M^2$.  The fractional added angular momentum would therefore be $\mathcal O(\mu)$. After a transient period in which fluctuations fall into the horizon or propagate away, the string will relax to rigid rotation with $\omega < \Omega_h$ and will begin angular momentum extraction.

If the CS is very long, energy will continually be drawn away from the BH and dissipated down the length of the string.  If the other end is sufficiently far away, in particular if the CS extends beyond the cosmological horizon, outward-moving fluctuations will never be reflected and return to the BH. As the string extracts angular momentum from the BH, its angular velocity will not increase significantly because that angular momentum propagates outward.  The BH will eventually spin down completely.

In the process of spinning down, the black hole can lose at most 29\% of its original mass.  For a typical, rapidly rotating, but nonextremal BH, the mass loss will 
be of the order of 10\%. Taking $\Delta M \approx M/10$ and ignoring the dependence of \eqref{eq:energy_extration_rate} on the BH spin, the timescale $t_\mathrm{SD}$ over which the black hole spins down is given by
\begin{equation}
\label{eq:spindowntime}
t_\mathrm{SD} \approx 10^{6} \, {\rm s} \left( \frac{M}{M_\odot} \right) \left(\frac{10^{-11}}{\mu}\right) \ .
\end{equation}
The fiducial tension $\mu = 10^{-11}$ is roughly the maximum value allowed by the observational bounds, implying the last factor will be greater than one.  Because the actual mass-loss rate goes to zero as $a \to 0$, the spin-down time given by \eqref{eq:spindowntime} is an overestimate.

For a stellar mass black hole with, say, $M_\mathrm{BH} = 30 M_\odot$, the approximate spin-down time \eqref{eq:spindowntime} gives
\begin{equation}
\label{eq:spindowntime_stellarmass}
t_\mathrm{SD} \approx 1 \, {\rm yr} \, \left(\frac{10^{-11}}{\mu}\right) \, .
\end{equation}
This implies that, unless $\mu$ is very small, a stellar mass black hole will spin down rapidly unless it is being actively  spun up by, for example, accretion.  In particular, a typical observation of a stellar BH will be extremely unlikely to see the spin down in progress.

In contrast, for a supermassive black hole, $M_\mathrm{BH} = 10^7 M_\odot$, \eqref{eq:spindowntime} gives
\begin{equation}
\label{eq:spindowntime_supermassive}
t_\mathrm{SD}  \approx 10^6 \, {\rm yr} \left(\frac{\mu}{10^{-11}}\right)^{-1} \ .
\end{equation}
The spin down of a supermassive BH is a much slower process, especially for small $\mu$. Consequently, the chance of observing such a BH in the process of spinning down will be much higher.

\section{Primordial black holes and cosmic strings}
\label{sec:PBH}

\subsection{Formation and spin up}
\label{sec:spin_up}

 Black holes may have been created in the early universe by a variety of mechanisms. The collapse of large density fluctuations during the radiation dominated era \cite{Zeldovich:1967lct,Carr:1974nx} has received the most attention,\footnote{Such scenarios require significant enhancement of cosmological perturbations at small scales.  This necessitates modifications to the standard slow-roll model constructed to match CMB observations, for example, by including non-minimal derivative couplings \cite{Heydari:2021gea}.} but other scenarios have been considered, such as the collapse of vacuum bubbles or domain walls \cite{Garriga:1992nm, Khlopov:2004sc,Tanahashi:2014sma, Garriga:2015fdk,Deng:2016vzba, Deng:2017uwc, Huang:2023chx, Huang:2023mwy} and of cosmic string loops or cusps \cite{Garriga:1992nm, Hawking:1987bn, Polnarev:1988dh}. These primordial black holes~(PBHs) can, in principle, be created at any mass within a very wide range.  Although the idea of PBHs is not new, the possibility that large black holes recently observed by LVK may have primordial origin \cite{Bird:2016dcv} has spurred renewed interest. In addition, recent work has argued that NANOGrav may have observed stochastic GW signals from PBH formation in the early universe, suggesting that PBHs comprise a significant fraction of dark matter~\cite{Kohri:2020qqd, DeLuca:2020agl, Vaskonen:2020lbd}.

The abundance of PBHs is constrained by a variety of phenomenological bounds \cite{Sasaki:2018dmp, Green:2020jor, Villanueva-Domingo:2021spv, Escriva:2022duf}. Extremely light PBHs, with $M_\mathrm{PBH} <  10^{-19} \, M_\odot$, would have completely evaporated by now due to Hawking radiation and would no longer be present. 
Somewhat more massive PBHs with $M_\mathrm{PBH} <  10^{-16} \, M_\odot$ would not have evaporated yet, but they would be emitting sufficient Hawking radiation that their abundance must be quite limited to have avoided observation. The abundance of PBHs in the planetary to subsolar-mass range, $10^{-11}\, M_\odot < M_\mathrm{PBH} < 10^{-1} \, M_\odot$, is modestly constrained by the frequency of stellar microlensing events, though for solar-mass and larger PBHs, $M_\mathrm{PBH} > M_\odot$, dynamical effects, limits on radiation from accreted gas, and GW signals impose more stringent limits.  The subplanetary mass range, $10^{-16}\, M_\odot < M_\mathrm{PBH} < 10^{-11} \, M_\odot$, however, is still quite unconstrained.

The mass at which a PBH forms is sensitively related to the time at which it forms. A region will collapse into a PBH of mass $M$ when its Schwarzschild radius is the size of the Hubble radius. During the radiation-dominated era, the average mass within a Hubble radius is $M_H = M_{\odot} \left(\frac{t}{10^{-5} \, \rm s}\right)$.
As a result, a PBH of mass $M$ will form at a time \cite{Vilenkin:2018zol, Villanueva-Domingo:2021spv}
\begin{equation}
\label{PBH_formation_time}
    t_\mathrm{PBH}  \sim 10^{-5} \, {\rm s} \left(\frac{M}{M_\odot}\right) \, .
\end{equation}

Unlike black holes formed by stellar collapse, PBHs formed by collapse of over-densities typically have very low spins.  Estimates in \cite{Mirbabayi:2019uph, DeLuca:2019buf} predict an initial spin parameter of order $a \sim 10^{-2}$.  Larger spins have been suggested, however, as a possibility for PBHs formed via other mechanisms \cite{Cotner:2017tir, Harada:2017fjm}.

Over the long history of the universe, both the mass and spin of a PBH can evolve significantly. While many phenomena can affect the PBH during its lifetime, accretion of the surrounding interstellar gas has been shown to be a significant driver of mass and spin evolution \cite{Ricotti:2007jk, Ricotti:2007au, DeLuca:2020bjf, DeLuca:2023tqx}.  In particular, the accretion effects grow with BH mass and become significant for $M \gtrsim 10 M_\odot$.  At early times, $z > 10$, the accretion rate for these large BHs can exceed the Eddington rate of $1.44 \times 10^{14} \, {\rm kg/s}\, (M/M_\odot)$.  For super-Eddington accretion rates, the accreting matter will form a thin disk \cite{Ricotti:2007au}, which leads to efficient spin-up of the PBH. 
If PBHs constitute only a fraction of the dark matter, the quantity of dark matter accreted is comparatively negligible.  However, the dark matter halo around the PBH can enhance the baryonic accretion rate by orders of magnitude \cite{Serpico:2020ehh, DeLuca:2023tqx}, enhancing the spin-up.

This rapid accretion process is expected to decrease with the onset of structure formation.  Due to the attraction of large-scale structures, a typical PBH will speed up relative to the ambient gas, and the resulting large proper velocities strongly suppress accretion \cite{Ricotti:2007au, Hutsi:2019hlw}.  Although uncertainties remain regarding the dynamics of PBHs and structure formation, significant accretion can be expected to end around $z \sim 10$.

The evolution of PBH mass and spin assuming a sharp accretion cut-off at $z=10$, was computed in \cite{DeLuca:2020bjf}, labeled as Model~1.  A small PBH with initial mass $M \lesssim 30 M_\odot$ and low spin $\chi \sim 0$ experiences negligible accretion and remains small and slowly spinning.  However, larger PBHs with $M > 30 M_\odot$ rapidly accrete as $z \to 10$, spinning up to near extremality $\chi \lesssim 1$.  In both cases, after accretion ends at $z=10$, PBHs are then assumed to continue with constant mass and spin until the present (see Fig. 4 of \cite{DeLuca:2020bjf}).

An alternative scenario, Model~2, was also considered in \cite{DeLuca:2020bjf}, in which the effect of structure formation is limited and accretion diminishes after $z=10$ but continues until the present. In this case, only very large PBHs, with $M > 10^3 M_\odot$, currently have high spins (see Fig. 5 of \cite{DeLuca:2020bjf}). The precise late-time spin of a high-mass PBH is therefore somewhat sensitive to the accretion history.   What matters for our argument, however, is only that, absent cosmic strings, large PBHs are driven toward high spin, a conclusion shared by both models.

The results of \cite{DeLuca:2020bjf} yield a prediction for the late-time spins of PBHs. In the absence of further dynamical evolution, a large PBH with mass above $M_\mathrm{min}$, where $M_\mathrm{min} > 30 M_\odot$ for Model~1 and $M_\mathrm{min} > 10^3 M_\odot$ for Model~2, and observed after $z=10$ would be expected to have near-extremal spin.  A small PBH, by contrast, should be observed with nearly zero spin.

As our subsequent discussion relies on this prediction, it is worth emphasizing there are still significant uncertainties in BH accretion in the early universe.  For example, chaotic instead of coherent accretion can reduce rather than increase the spin.  See \cite{Regan:2024wsu} for a recent review of BH accretion.

\subsection{Spin down by cosmic strings}
\label{sec:spin_down}

The presence of cosmic strings changes the predicted distribution of PBH spins significantly. 
Cosmic strings form early in the history of the universe, generally before the formation of PBHs. However, in order to be relevant to PBH, they must have formed after inflation; otherwise, the string network at the time of PBH formation would be much too sparse.  Focusing on cosmic strings created during symmetry-breaking phase transitions, the energy scale $\eta$ of the phase transition, which is related to the string tension $\mu = \left(\frac{m_p}{\eta}\right)^{2}$, determines the time of string formation \cite{Vilenkin:2018zol},
\begin{equation}
    t_s \sim \left(\frac{m_p}{\eta}\right)^{2} t_p \sim \frac{ t_p}{\mu}
\end{equation}
where $t_p$ is the Planck time. Compared with the PBH formation time \eqref{PBH_formation_time}, cosmic strings form much earlier than PBHs, as long as the CS tension $\mu$ is not extremely small; for example, for a $30 M_\odot$ PBH any tension $\mu > 10^{-38}$ will yield $t_s \ll t_\mathrm{PBH}$.

After the strings are created, the network begins to evolve.  Long strings spanning across Hubble volumes stretch as the universe expands.  Through self-intersections and intersections with other strings they can reconnect and break off smaller loops.  Loops oscillate, lose energy by emitting gravitational waves, and shrink.  Numerical simulations predict a scale-invariant string network, in which a given Hubble patch contains several long strings and a larger number of loops ranging over all sizes \cite{Bennett:1989, Allen:1990tv, Blanco-Pillado:2011egf}.

Later, after this CS network has emerged, PBHs begin to form. Overdense regions begin to collapse when their apparent horizon comes within the cosmological horizon \eqref{PBH_formation_time}.  This roughly cosmological-horizon-sized region, which includes $\mathcal O(10)$ long strings that cross through it, then collapses, forming a new PBH.  For every long CS originally in the collapsing region, the new PBH will have a pair of long strings sticking out of the horizon \cite{Vilenkin:2018zol}.\footnote{However, alternative PBH formation scenarios, such as collapsing bubbles, in which the black hole horizon is much smaller than the cosmological horizon, will result in few, if any, captured strings.}

The subsequent evolution of the coupled PBH-CS system has been studied in detail in Ref.~\cite{Vilenkin:2018zol, Xing_2021, Deng:2023cwh}.  A long, horizon-crossing string attached to a black hole horizon exerts a pull on the black hole but cannot, by itself, become unattached from the black hole.  If, for example, a string connects two black holes, they can be pulled toward each other, eventually merging into a single larger black hole with no cosmic string.  However, if the typical distance between PBHs is large, the string pulling the PBHs together will have to compete with the cosmological expansion.

A pair of strings emerging from a BH horizon can intersect, potentially reconnecting to form a loop attached to the horizon and a long string, which is disconnected from the PBH.  The half loop may eventually fall into the horizon and disappear, leaving the PBH with two fewer strings. Initially, with $\mathcal O(10)$ strings attached to the PBH, the probability for any two to intersect is high, but, as the number of strings decreases, intersections become increasingly unlikely~\cite{Vilenkin:2018zol}.  

The details of the string dynamics are complicated and depend on the properties of the cosmic strings, the density of the CS network, and the distance between PBHs. The likelihood that a PBH will eventually lose all its strings, and the timescale for this to occur, are open questions and will likely require numerical simulations such as those of Ref.~\cite{Deng:2023cwh}.  We will parameterize this uncertainty by a single survival probability $P_\mathrm{surv}$, defined as the probability that at least one long, horizon-crossing string remains attached to the PBH throughout the spin-down time $t_\mathrm{SD}$.  Because a newly formed PBH carries $\mathcal{O}(10)$ attached strings and intersections become increasingly unlikely as the number of strings falls \cite{Vilenkin:2018zol}, $P_\mathrm{surv}$ is not expected to be parametrically small, although a more precise estimate will require numerical simulation \cite{Deng:2023cwh}.

Although we focus here on energy extraction due to long cosmic strings, it was shown in Refs.~\cite{Xing_2021, Deng:2023cwh} that a short string loop with both ends attached to the horizon can also extract rotational energy from a spinning BH.  Rather than falling into the horizon, if the BH spin is sufficiently large, the string will become a rotating ``double-line," which becomes increasingly elongated, thereby extracting the rotational energy of the BH.

As described above in Sec.~\ref{sec:spin_up}, a typical PBH forms with very low spin, but large PBHs will subsequently be spun up by accretion. However, assuming the large PBH has one or more cosmic strings attached during the accretion period, the angular momentum will then be extracted by the cosmic strings, spinning the PBH back down.

The spin-down rate for a $30 M_\odot$ mass PBH \eqref{eq:spindowntime_stellarmass} is quite rapid, as long as the cosmic string tension is not extremely small.  Even for $\mu = 10^{-20}$, orders of magnitude below current observational bounds, the spin-down time is $10^{9}$ years; by the present day, the PBH will have long since lost its angular momentum.  For larger string tensions, closer to the observational bound $\mu \sim 10^{-11}$, the spin-down rate is faster than the accretion rate \cite{DeLuca:2020bjf}.  In this case, the PBH never even gets spun up, with the angular momentum extracted as fast as it falls in.

To make this argument more explicit, we can estimate the statistics of observed PBH spins.  Consider the population of high-mass PBHs, with $M > M_\mathrm{min}$, which in the absence of cosmic strings, would be observed with near-extremal spin \cite{DeLuca:2020bjf}.  The fraction $f_0$ of this population that is instead observed with near-zero spin can be written schematically as
\begin{equation}
\label{eq:lowspin_fraction}
f_0 \;\simeq\; P_\mathrm{attach}\,\times\, P_\mathrm{surv}\,\times\, \Theta(t_\mathrm{SD} < t_\mathrm{obs}) \, ,
\end{equation}
where $P_\mathrm{attach}$ is the probability that the PBH forms with at least one long string attached, $P_\mathrm{surv}$ is the survival probability defined above, and the step function $\Theta$ enforces that spin-down completes before the time $t_\mathrm{obs}$ when the observation is made.  Two of these three factors are favorable essentially by construction.

For the attachment probability $P_\mathrm{attach}$, a PBH forms within a horizon-sized region already threaded by $\mathcal{O}(10)$ long strings, so $P_\mathrm{attach} \approx 1$ \cite{Vilenkin:2018zol}.\footnote{This is not the situation in the supermassive BHs case discussed in Sec.~\ref{sec:supermassiveBH} below, for which the probability of attachment is significantly lower.}  Likewise, the spin-down condition $t_\mathrm{SD} < t_\mathrm{obs}$ is satisfied for any $\mu > 10^{-20}$, as shown by \eqref{eq:spindowntime_stellarmass}, so $\Theta = 1$ across the entire tension range of interest.  The prediction is therefore controlled mostly by the survival probability, $f_0 \simeq P_\mathrm{surv}$.  Even a conservative survival probability, $P_\mathrm{surv} \sim \mathcal{O}(0.1)$, would produce a substantial and statistically distinguishable depletion of high-spin objects among high-mass PBHs relative to the near-extremal expectation of \cite{DeLuca:2020bjf}.  A more complete treatment, incorporating the collision and capture statistics, the velocity and orientation distributions at capture, and the non-steady initial phase, lies beyond the rigidly-rotating approximation adopted here and is left to future numerical work.  A more accurate calculation of $P_\mathrm{surv}$ will rely on a detailed understanding of the dynamics of attached strings and numerical simulations \cite{Deng:2023cwh} going beyond the approximations of this work.

This allows us to establish a clear prediction and a definitive test for the existence of cosmic strings in an unprobed range of string tensions. If there are cosmic strings with $\mu > 10^{-20}$,  PBHs with mass larger than $30 M_\odot$ will be observed to have near-zero spin.  Otherwise, according to \cite{DeLuca:2020bjf}, these high mass PBHs will instead have near-extremal spin.

For string tensions near the observational bound, $\mu \sim 10^{-11}$, the spin-down rate exceeds the accretion-driven spin-up rate, so an accreting PBH never reaches high spin; in this regime, the prediction of near-zero spin for large-mass PBHs is independent of which accretion model of \cite{DeLuca:2020bjf} is adopted.

This conclusion relies on several assumptions and approximations, which we summarize as follows:
\begin{itemize}
    \item For a PBH, $P_\mathrm{attach} \approx 1$ at the time of formation due to strings threading the collapsing region. \
    \item  At least one long, horizon-crossing string survives for a time $t_\mathrm{SD}$, parametrized by $P_\mathrm{surv}$. \
    \item  After an initial transient period, the system relaxes to a string rigidly rotating with $\omega < \Omega_h$.\
    \item  The cosmic string tension is $\mu > 10^{-20}$, so the spin-down is sufficiently fast: $t_\mathrm{SD} < t_\mathrm{obs}$. \
    \item Accretion consistent with \cite{DeLuca:2020bjf} Model~1 or Model~2, such that in the absence of attached cosmic strings, large PBHs are spun up to near extremality.  The minimum mass is model dependent, ranging from $M_\mathrm{min} > 30 M_\odot$ for Model~1 and $M_\mathrm{min} > 10^3 M_\odot$ for Model~2.
\end{itemize}

\subsection{Observations of black hole spin}
\label{sec:Observations}

Determining the spin of a BH presents a significant observational challenge. Currently, spin measurements are possible for two types of BH systems: Spins of accreting BHs can be measured through electromagnetic radiation, mostly through X-ray reflection spectroscopy, and thermal continuum fitting, while spins of merging black holes are constrained through GW observation (see \cite{Reynolds:2020jwt} for a review).

For accreting BHs in the regime between geometrically thin and optically thick accretion disks, X-ray reflection spectroscopy is the primary means for determining the spin. X-ray reflection spectroscopy refers to the soft X-ray lines emitted by the photo-ionized outer region of the optically thick accretion disk. The applicable regime corresponds to accretion rates between about 0.01 and 0.3 of the Eddington accretion rate and, in some circumstances, is applicable up to about the Eddington limit. This technique is applicable to classical Seyfert galaxies, moderately luminous quasars, and black hole X-ray binaries in their luminous hard X-ray state.

Thermal continuum fitting utilizes the fact that the spin of an accreting BH influences the location of the innermost stable circular orbit (ISCO) in the accretion disk, thus determining the temperature of the blackbody radiation from that region. For prograde accretion, the higher the spin rate of the BH, the smaller the ISCO and the higher the temperature of the inner disk. This technique has been applied more to stellar-mass BH X-ray binaries than to accreting supermassive BHs.

Using these methods, many accreting supermassive BHs have been found to be rapidly spinning. However, there is a population of more slowly spinning BHs with masses greater than $3\times 10^7 M_\odot$, consistent with what is expected from structure formation models. Accreting stellar-mass BHs in X-ray binary systems have also been observed to be rapidly spinning, but, in many cases, their spins must have been rapid at birth instead of having been produced by spin-up accretion. Spin-up from near-zero spin to maximal spin requires that a significant fraction of the BH mass be provided by accretion. In the case of high-mass X-ray binaries, the accretion process is time-limited due to the rapid evolution of the massive companion star. Even Eddington-limited cases would not have sufficient time to spin up the BH by accretion. For low-mass X-ray binaries, however, spin up by accretion is possible.

Likely more relevant to the PBHs being discussed here are measurements of the spins of merging binary black holes (BBH) using the GW signal. These observations are cleaner because details of the accretion do not come into play. However, the effects of spin in the GW signal are subtle, and the sensitivity of LVK is such that, for the cases observed, the results presented in \cite{Reynolds:2020jwt} interestingly favor low spins for the merging objects.

Beyond measuring a black hole's spin, there remains the task of identifying whether it is primordial in origin.  BBH merger events observed in the LVK third GW Transient Catalog (GWTC-3) \cite{KAGRA:2021vkt} have widely been considered as a mixed population of astrophysical black holes (ABHs) and PBHs \cite{Hall:2020daa, Hutsi:2020sol, Chen:2024dxh}. There are aspects of these observations that suggest the existence of a subpopulation of PBHs. This includes the merger event rate of high mass BBH systems, including events that reside in the upper mass gap that applies to ABHs \cite{LIGOScientific:2020iuh}. A recent Bayesian analysis estimated that $1/4$ of the events in the GWTC-3 are due to PBHs \cite{Chen:2024dxh}. However, because the mass spectrum of PBHs is expected to peak at high mass ($30 M_\odot$) \cite{Hutsi:2020sol}, observations of high-mass BBH merger events are significantly more likely to be PBH systems. These systems are the ideal candidates for the observation described here because of the spin-up process described above. Repeated observations of such high-mass merger events will allow for further delineation of sub-populations of BBH mergers and would allow for a definitive test for the existence of cosmic strings with a string tension $\mu > 10^{-20}$.

\section{Spin-down of Supermassive Black Holes}
\label{sec:supermassiveBH}

As discussed in Sec.~\ref{sec:PBH}, primordial black holes form with cosmic strings preattached.  However, later-forming, non-primordial black holes can only become attached to a CS if the two  collide.  A stellar mass BH is an extremely small target, and the probability for a passing cosmic string to hit it is vanishingly low.  A supermassive BH, on the other hand, is large enough that collisions with cosmic strings, although still rare, can be expected to occur over the lifetime of the BH in certain circumstances.

The rate at which strings collide with and become captured by a supermassive BH depends on size of the BH and the density of the CS network. Cosmic strings tend to cluster in galactic halos \cite{Jain:2020dct} where the BHs are located, improving the chances for a collision. The expected time for a CS collision was estimated in \cite{Xing_2021} as
\begin{equation}
\label{eq:capture_time}
t_c = 3 \times 10^{9}\, \rm{yr} \left(\frac{\mu}{10^{-18}}\right)^{2} \left(\frac{M}{M_\mathrm{SgrA*}}\right)^{-2.5}\left(\eta(p)\right)^{-1}
\end{equation}
where the mass of Sgr A*, $M_\mathrm{SgrA*} = 4 \times 10^6 \ M_{\odot}$, is taken as a fiducial mass, and $\eta(p)$ is an order-one function of the CS reconnection probability $p$. In the particular case of Sgr A*, if the cosmic strings are rather light, $\mu < 10^{-18}$, we would expect at least one CS to have been captured since the BH's formation, order $10^{10}$ years ago.\footnote{Although \eqref{eq:capture_time} was derived for supermassive black holes, if we extrapolate the expression down to a stellar BH mass, $M_\mathrm{BH} \sim 10 \ M_{\odot}$, the capture time becomes far longer than the age of the universe, unless the tension is extremely small: $\mu < 10^{-46}$. Consequently, the probability that a BH formed from stellar core collapse ever captures a CS is negligible.}

If a supermassive BH were to capture a CS, either a long string or a string loop, there would be observational consequences. For example, gravitational lensing could be used to detect the acceleration of a BH caused by the net force exerted by attached cosmic strings \cite{Ashoorioon:2022zgu}. Oscillating or rotating cosmic strings, extracting rotational energy from the BH, could emit observable gravitational waves, although the particular signatures from such a process are still to be determined \cite{Xing_2021}. We propose an alternative method to observe a long cosmic string spinning down the BH.

Supermassive BHs are expected to be rapidly spinning, and those that have been observed all have large spins \cite{Reynolds:2020jwt}.  Lower mass supermassive BHs, those with $M < 3 \times 10^7 M_\odot$, all have near-extremal spins, $\chi > 0.9$.  The heaviest supermassive BHs, on the other hand, tend to have more modest spins in line with the cosmological simulations \cite{Sala:2023uxf}.

As shown in Sec.~\ref{sec:energy_extraction}, once a CS has been captured by the BH, it will begin extracting angular momentum and, along with it, energy. This spin-down process takes a long time \eqref{eq:spindowntime_supermassive} because there is a lot of angular momentum to extract.  In the case of Sgr A*,  the spin-down time would be $t_\mathrm{SD}  > 10^9 \, (\mu / 10^{-14})^{-1} \, {\rm yr}$, implying that,  for $\mu < 10^{-14}$, however long ago the string was captured, the spin down would be currently ongoing.

In principle, the decreasing BH spin could be measured using the methods discussed in Sec.~\ref{sec:Observations}, for example, using X-ray reflection spectroscopy for an accreting supermassive BH.
However, the spin is only decreasing at a rate of $\dot a/a \sim t_{SD}^{-1} \sim 10^{-9} \, (\mu / 10^{-14}) \, {\rm yr^{-1}}$ for Sgr A*, and the accuracy of these techniques is not nearly good enough to measure the spin to one part in $10^9$ \cite{Reynolds:2020jwt}.

Although the BH's spin is challenging to measure, its mass $M$ can be readily determined from observations of orbiting objects. In particular, a pulsar orbiting the BH would allow a very accurate measurement of the mass.  Pulsars are the most accurate clocks in nature and have long played an important role in high-precision astrophysical measurements.  Famously, the gradually decreasing period of the Hulse-Taylor binary pulsar was measured to one part in $10^{15}$, providing the first observational evidence for gravitational waves \cite{Taylor:1982,Weisberg:2010zz}.

Supermassive BHs are found in dense galactic centers, which are also expected to be the home of large numbers of compact objects.  In our own galaxy, around 100 pulsars are predicted to be orbiting very close to Sgr A*, with orbital periods $P \lesssim 10 \, \rm yr$~\cite{Pfahl:2003tf, Zhang:2014kva, Bartels:2015aea}. The Square Kilometer Array (SKA) is expected to probe the galactic center and observe these pulsars.

Assuming such a pulsar is in fact found close to Sgr A*, timing observations will be able to very accurately measure $P$.  Ongoing observations over an entire period will then determine the rate at which the period is changing.  If Sgr A* is losing mass due to extraction by cosmic strings, its gravitational attraction on the pulsar will decrease, causing the pulsar to spiral outward and the period to increase\footnote{In principle, there is a competing effect due to gravitational wave emission of the orbiting pulsar which causes it to inspiral and for the period to decrease. However, this leads to a change in the period, which is very small, far below the level which can be detected. For $M_\mathrm{BH} = 10^6 M_\odot$ and $P = 10 \, \rm yr$, the rate of inspiral is approximately  $\dot P = 10^{-15}$ \cite{Simonetti:2010mk}.}~\cite{Hadjidemetriou1,Hadjidemetriou2,Simonetti:2010mk},
\begin{equation}
\label{eq:period_increase_rate_general}
\frac{\dot P}{P} =-2 \frac{\dot M_\mathrm{SgrA*}}{M_\mathrm{SgrA*}} \ ,
\end{equation}
where the mass loss rate $\dot M$ is given by \eqref{eq:mass_loss_rate}.  For a pulsar with period $P = 10 \, \rm yr$, the period would be increasing at a rate of $\dot P \sim 10^{6}\, \mu$.  Only sufficiently light strings with $\mu < 10^{-18}$, would have a significant chance of being captured by Sgr A*, implying the effect on the period would be quite small:
\begin{equation}
\label{eq:period_increase_rate}
\dot P \lesssim 10^{-12} \ .
\end{equation}

Would such a small effect be observable? With current observational technology, the orbital period of an isolated pulsar orbiting near Sgr A* could, in principle, be measurable to an accuracy of about $\Delta P \sim 10^{-11} \ \rm yr$ \cite{Kovacs:1981, Larsson:1996, Simonetti:2020ivl}.  Using two measurements spaced apart by one period, the rate of change of the period could then be measured to an accuracy of $\dot P \sim \Delta P/P \sim 10^{-12}$, which is within the order of magnitude of the size of the expected effect \eqref{eq:period_increase_rate}.  
However, the neighborhood around Sgr A* is rather crowded, and a pulsar in this region would likely not be sufficiently isolated to achieve this precision. Scatter broadening of the radio pulses may also play a more significant role for observations toward the galactic center, which would require extended observations to beat down the uncertainty in determinations of pulse times of arrival; observations at higher frequencies mitigate the effects of scattering. The scattering along various lines of sight can vary quite a bit, so it is not easy to estimate how much effect scattering will have until a useful pulsar is found. Although the mass loss is, in principle, observable via pulsar measurements, significant observational improvements would be required to make it viable in practice. With the ever-improving observational facilities in astronomy, we expect that we are on the verge of making these observations ever more possible, but predicting future improvements or the rate of those improvements is not simple.

In addition, other factors complicate this proposed observational approach.  Sgr A* is continually accreting matter from its surroundings, and the resulting increase to the BH mass would be both large and uneven. However, a pulsar orbiting well outside any material surrounding the BH would measure, via Kepler's 3rd law, the combined mass of both the BH and the accreting matter. The gravitational effect on the pulsar of matter in an accretion disk does not change once it falls into the BH.  In contrast, a long CS extracts the mass far beyond the pulsar orbit and so reduces the gravitational pull on the pulsar and increases the orbital period.

One might be concerned that other environmental effects could mimic or obscure the pulsar timing observations we suggest. It is worth noting that as the BH loses mass, any orbiting pulsar will spiral outward, the orbital period increasing in time systematically. This is the opposite effect caused by other systematic environmental perturbations on the BH-pulsar system. Drag on the pulsar by gas, for example, will decrease the energy of the system, causing the orbital period to decrease. For a closely orbiting pulsar (the easiest case for performing human time-scale observations), the gravitational interplay between the BH-pulsar system and more distant stars would act to spread energy among the various entities (a result of the second law of thermodynamics, if you wish), resulting in a long-time decrease in the orbital period of the pulsar, systematically, albeit accompanied by random fluctuations. Long-term observations would be necessary to remove the random effects. 

It is instructive to consider the magnitude of the effect of drag by gas on the pulsar. Note the fractional change in orbital period is $\Delta P/P = -3\Delta E/2E$, where $E$ is the total orbital energy of the BH-pulsar system. At most, each gas particle in the medium would acquire twice the velocity of the passing neutron star (by an elastic collision), and the total energy removed from the BH-pulsar system would just depend on the mass of gas swept out by the neutron star, in its orbit. Thus, the fractional change in period is 
\begin{equation}
\label{eq:period_change_drag}
\frac{\dot P}{P} \sim -\frac{m_H n \pi R_{NS}^2 2 \pi r}{M_{NS}}, 
\end{equation}
where $m_H$ is the hydrogen mass, $n$ is the environmental number density of hydrogen atoms, $R_{NS}$ is a neutron star radius, $M_{NS}$ is a neutron star/pulsar mass, and $r$ is the orbital radius. For a SMBH, a 10-year orbital period, and $n\sim 1$~cm$^{-3}$, $\dot P
/P \sim 10^{-47}$ --- drag is a negligible issue.

As in the PBH case discussed in Sec.~\ref{sec:spin_down}, long, horizon-crossing strings can effectively carry away the extracted energy and angular momentum and spin down the BH.  A small CS loop captured by the BH would not spin it down using the same mechanism, as the energy could just travel all the way around the loop and fall back in. However, as discussed in Refs.~\cite{Xing_2021, Deng:2023cwh}, an attached string loop may evolve into a co-rotating and elongating ``double-line" which will also extract rotational energy and spin down the black hole.

\section{Discussion and open questions}
\label{sec:discussion} 

Many physical scenarios for the early universe predict cosmic strings, and, indeed, F- or D-strings of string theory could manifest as cosmic strings. Despite their ubiquity in such theories, scant evidence of their actual existence has been found.  A very intriguing statistical gravitational wave background signal was found recently by NANOGrav consistent with cosmic strings \cite{Ellis:2023tsl}. However this GW background turns out, alternative methods of testing for the existence of cosmic strings are quite valuable.  Here, we have considered the resulting spin-down of a black hole due to long, attached cosmic strings, for various black hole sizes, finding something of an inverse ``goldilocks'' problem in which the extremes may be preferred avenues for observation.

The prospects of observing the spin-down of stellar-mass black holes (with masses up to hundreds of solar masses) are dim for two reasons. First, the likelihood of such a black hole encountering a cosmic string is quite small, and, second, the spin down of the black hole would be quite fast and unlikely to be observed.

A supermassive black hole (of order a million or more solar masses), on the other hand, suffers from neither of these problems. Its size makes for a significant probability of encountering a cosmic string and for a long spin-down time. Because the spin down is very gradual, observing it requires extremely sensitive measurements of the black hole spin.  We suggest observing a pulsar orbiting close to Sgr A* and looking for out-spiraling due to energy being extracted from the black hole by the cosmic string.  However, the signal is likely too small even for high-precision pulsar observations, at least with current technology.

Primordial black holes, however, are likely to form with one or more cosmic strings already attached.  Although they typically don't form  with significant spin, PBHs soon get spun up by accretion, and, in the absence of cosmic strings to spin them down, large PBHs with masses over $30 M_\odot$ are predicted to reach near-extremal spins.  This allows a sharp test for the existence of cosmic strings: Large PBHs should be observed with almost no spin if cosmic strings exist and with very high spins if they do not.

This argument relies on some important assumptions and approximations, which were summarized at the end of Sec.~\ref{sec:spin_down}.  Several additional factors also complicate the interpretation offered above which we have largely omitted from our analysis but which could potentially play significant roles.

We have presented a simplified description of the dynamics of the black hole-cosmic string system.  For the purposes of estimating the energy extraction rate, we have ignored the initial, transient dynamics and focused on late-time quasi-stationary state in which the strings can be treated as rigidly rotating.  Multiple long strings attached to a BH can interact with each other, for example, by intersecting and reconnecting, resulting in a short segment with both ends attached and which falls into the BH and another segment no longer attached to the BH. Other than the energy and spin extraction, other effects of the string on the black hole have been neglected.  For example, an attached string would exert a force on the black hole, causing it to move. A more thorough discussion of these dynamics can be found elsewhere \cite{Vilenkin:2018zol, Xing_2021, Deng:2023cwh}.

We have also neglected other aspects of black hole evolution, in particular, merger events. PBHs may tend to form in clusters \cite{Choi:2025eqn, Zhang:2026vjk}, implying that mergers may be especially significant.  Typically, BH mergers result in highly spinning remnants, affecting, for example, the distribution of masses and spins of the BH population.  As a particularly interesting example, two PBHs connected by a cosmic string would get pulled toward each other and eventually merge.  If other strings remained attached, the resulting remnant could then be spun down.  Such a scenario merits future investigation.  More discussion of PBH evolution can be found in \cite{Hutsi:2019hlw} and \cite{DeLuca:2020bjf}.\footnote{See especially, the Appendix of \cite{DeLuca:2020bjf} for discussion of PBH mergers.}

Finally, differentiating PBHs from conventional stellar-sized BHs presents an interesting statistical challenge (see Ref.~\cite{DeLuca:2020bjf}), especially given that future generations of GW detectors will greatly expand the volume range of BH observations and that merger remnant BHs will be highly spinning.

\acknowledgments
We would like to thank Edward Smakov, Michael Ramsey, and Timmy Dhakaia who were involved in early stages of this project.  MK would like to thank J.~Kanner, J.~Papa, P.~Shawhan, and T.~Vachaspati for their insightful discussions.
The work was supported by the National Science Foundation grants PHY-1820733, PHY-2000398, PHY-2310608 (ML); AST-2011731 (MK); PHY-2011383, PHY-2308861 (SL); and AST-2011757 (JS). 
ML would also like to thank both the APCTP and the BNL EIC Theory Institute for their warm hospitality while this work was being completed.

\bibliographystyle{JHEP}
\bibliography{newbib}

\end{document}